\documentclass[12pt,a4paper]{article}

\usepackage[cp1251]{inputenc} 
\usepackage[T1]{fontenc} 
\usepackage{microtype} 
\usepackage{mathrsfs}
\usepackage[english]{babel} 
\usepackage{bbm}
\usepackage{amsmath,amsfonts,amscd,amssymb,latexsym}
\usepackage{dcolumn}

\usepackage[unicode, pdftex]{hyperref}
\usepackage[usenames]{color}

\usepackage{graphics}
\usepackage{titling}
\usepackage{ textcomp }
\graphicspath{{/}}
\DeclareGraphicsExtensions{.eps,.pdf,.png,.jpg}
\usepackage{wrapfig}
\usepackage{amssymb}
\usepackage{amsmath}
\usepackage[dvips]{graphicx}
\usepackage{rotating}
\usepackage{longtable}
\usepackage{multirow}
\usepackage{geometry}
\begin{document}


\newenvironment{sciabstract}{
\begin{quote} \scriptsize}
{\end{quote}}

\pretitle{\begin{center}\Huge\bfseries} 

\title{
Quarks in Hadrons and Nuclei\\
\Large{Part 1: Quark Structure of Nucleon}}

\author
{G. Musulmanbekov$^{*,1,2}$\\
\scriptsize{$^1$JINR, Dubna, Russia}\\
\scriptsize{$^2$INP, Almaty, Kazakhstan}\\
\scriptsize{$^*$e-mail: genis@jinr.ru}\\
}

\date{}
\maketitle


\begin{abstract}
We propose a semi-empirical quark model of nucleon structure, so-called, Strongly Correlated Quark Model, SCQM, which possess the features of both non-relativistic and relativistic quark models. Based on SU(3) color symmetry it includes the main features of QCD: local gauge invariance, asymptotic freedom, and chiral symmetry breaking.
This First Part of the paper is devoted to description of the model, SCQM, and in the forthcoming Second Part we will apply the model to built the nuclear structure. Applied to nuclei, it reveals emergence of the face-centered cubic (FCC) symmetry of the nuclear structure.This symmetry arise from quark-quark correlations leading, in turn, to nucleon-nucleon correlations.
\vspace*{6pt}

\noindent

\end{abstract}

\section*{Introduction}

One of the major goals of nuclear physics is to provide the information how neutrons and protons stick together forming the nuclear structure. Historically there are three  well known conventional nuclear models based on different assumption about the phase state of the nucleus: the liquid-drop, shell (independent particle), and cluster models. The liquid-drop model requires a dense liquid nuclear interior (short mean-free-path, local nucleon interactions and space-occupying nucleons) in order to predict nuclear binding energies, radii, collective oscillations, etc. In contrast, in the shell model each point nucleon moves in mean-field potential created by other nucleons; the model predicts nuclear spin and parity, the existence of nucleon orbitals and shell-like orbital-filling. The cluster models require the assumption of strong local-clustering of particularly the 4-nucleon alpha-particle in order to make predictions about the ground and excited states of cluster configurations. The dilemma of nuclear structure theory is that these mutually exclusive models work surprisingly well for qualitative and quantitative explanation of certain limited data sets, but each model is utterly inappropriate for application to other data sets. There is a wide variety of attempts (on the whole unsuccessful) to solve the problem of nuclear structure employing realistic nucleon--nucleon interactions. Even three-body forces, introduced to improve the situation, do not provide a solution.

The fundamental theory of the strong interactions, Quantum Chromodynamics, QCD, \cite{QCD}, is applicable in the high-energy regime, i.e., derivation of dynamical structure of hadrons and, especially, nuclei from the first principles of QCD remains the unsolved problem. Quark degrees of freedom manifest themselves at high-momentum transfers in lepton/hadron--hadron and lepton/hadron--nucleus interactions. For the most low-energy applications, solutions of the QCD Lagrangian can only be obtained numerically with methods of lattice QCD. However, the lattice QCD does not specify the overall dynamics of the system; according to the Lagrangian density of QCD quarks and gluons interact with each other in a very complicated way. Hence, the most important problem of nuclear physics concerns the role of quarks in organizing the nuclear structure: how are nucleons bound inside nuclei and do quarks manifest themselves explicitly in ground-state nuclei? Nowadays very popular tools in theoretical nuclear physics are effective field theories, EFT \cite{van Kolck}.
The main idea of EFTs based on QCD is to expand the Lagrangian density of QCD in terms of a typical momentum for the process under consideration. It is realized by replacing the quark and gluon fields of QCD by a set of meson fields which are the Goldstone bosons of the theory due to the dynamical breaking of chiral symmetry.

However, the fundamental question of nuclear physics is still remaining: Do quarks play an explicit role of arranging nucleons in nuclear structure? This is the aim of our paper. We argue that nucleons within nuclei are bound with each other via color charged quark--quark interactions which lead to strong nucleon--nucleon correlations. This aim is realized in the framework of the, so-called, Strongly Correlated Quark Model (SCQM) \cite{Mus1, Mus2, Mus3} of hadron structure, which is presented in this Part of the article. In Part 2 the SCQM will then be applied to build the nuclear structure. We will demonstrate that the emerging geometry of nuclear structure corresponds to symmetry of the face-centered cubic (FCC) lattice.  The model is isomorphic to the shell model and, moreover, composes the features of cluster models.

\label{sec:preparation}
\section{Ingredients of QCD}
\label{sec-1}
 According to QCD, nucleons are composed of three valence quarks, gluon fields and sea of quark-antiquark pairs. Quarks possess various quantum numbers: flavour (u, d), electric charge (+2/3, -1/3), spin (1/2) and color (Red, Green, Blue). Exchange particles mediating interactions between quarks are ``gluons'' possessing spin 1 and different colors:
\begin{equation}
 R\bar{G}, G\bar{R}, R\bar{B}, B\bar{R}, G\bar{B}, B\bar{G}, R\bar{R}, G\bar{G}, B\bar{B}
\label{9gluons}
\end{equation}
From latter three gluons one can make two combinations
\begin{equation}
  \sqrt{\frac{1}{2}} (R\bar{R}-G\bar{G}), {\sqrt{\frac{1}{6}} (R\bar{R}+G\bar{G}-2B\bar{B})}.
\label{2gluon}
\end{equation}
As a result there are eight linearly-independent combinations of gluons mediating interactions between quarks. However, derivation of nucleon properties from the first principles of QCD is still not the solved task. Hence, there is a variety of phenomenological models which can be united in two groups: current quark models and constituent quark models. In current quark models relativistic quarks having masses 5--10 MeV move freely inside the restricted volume or a bag. In constituent quark models the quark-antiquark in mesons and three quarks in baryons are non-relativistic, surrounded by quark-antiquark sea and gluon field.
 Although the quark and qluon dynamics is described explicitly by the Lagrangian density of QCD we don't know how quarks and gluons are distributed inside the nucleon. In order to learn something more about the structure of nucleon, the elaborated models  should involve one or more general features of QCD. There are three general ingredients of the theory: first, QCD possesses {\it local gauge invariance}; second, it is {\it asymptotically free}; third, low energy approximation of QCD exhibits {\it chiral symmetry breaking} \cite{ChSB}.

What is chiral symmetry? QCD lagrangian for 2 flavors ($u, d$) is known to possess a
symmetry under U(2)$_{\text{L}}\times $ U(2)$_{\text{R \ }}$ independent
rotations of left--handed and right--handed quark fields which are two
component Weyl spinors. This symmetry is called \textit{chiral.}
 We know that this symmetry is broken. The order parameter for
symmetry breaking is the quark or \textit{chiral condensate,}

\begin{equation}
\left\langle \overset{-}{\psi }\psi \right\rangle \simeq -(250\text{ }%
MeV)^{3},\text{ \ \ \ \ \ \ \ \ \ \ \ }\psi =u,d
\end{equation}

As a consequence of chiral symmetry breaking massless valence quark (u, d) acquies dynamical mass which we call constituent quark. Its value at zero
momentum can be estimated as one half of the $\rho $ meson mass or near one third
of the nucleon mass
\begin{equation}
M_{Q}\simeq 350-400\text{ }MeV.
\end{equation}
There are different approaches to describe the mechanism of chiral symmetry breaking ( instanton -- induced interactions, monopole condensation, etc.).
According to the SCQM, as shown below, chiral symmetry breaking is reaction of the physical vacuum on presence of a single color quark, and a dynamical mass of quarks originates from destructive interference of color fields of quarks.
\section{\bf Strongly Correlated Quark Model~}
\label{sec-2}
\subsection{Model Ingredients}
One of the aims in physics is not only to calculate some observables and get
a correct numbers but mainly understand the physical picture for given
phenomenon. It very often happens that a theory formulated in terms of
fundamental degrees of freedom cannot answer such a question since it
becomes overcomplicated at the related scale. We believe that underlying theory of strong interactions is QCD. However, the
derivation of the features of hadrons and nuclei from the first principles
of QCD still remains a big problem. There are two types of phenomenological
models of hadron structure: non-relativistic (constituent) quark models, NRQM, which work
well in hadron spectroscopy and soft processes \cite{NRQM}, and relativistic (current)
quark models which are exploited for description of DIS and hard processes. The least example of this class of models is MIT bag model \cite{MIT} where massless quarks are confined in a bag. Although we believe that constituent quarks are composed of valence quarks dressed by gluons and ''sea'' quark--antiquark pairs, the relationship
between relativistic and non-relativistic models still remains smooth and unclear. Now it is believed that this relationship could be formulated as chiral symmetry breaking of QCD.
To describe the nucleon representing essential features of QCD we elaborated the Strongly Correlated Quark Model, SCQM. Although the model is rather qualitative than quantitative, it allows to construct in a semi-empirical way the nuclear structure starting from the lightest nuclei through medium towards heavy ones. The ingredients of the model are the following.
The physical vacuum, whose energy is below the ``empty'' perturbative vacuum,  is populated by gluon and quark--antiquark condensate. Imagine a single quark of a certain color embedded in the physical vacuum. The color field of the quark polarizes the surrounding vacuum creating condensate.  At the same time it experiences the pressure of the vacuum, as a reaction on ordering, because of the presence of quantum fluctuations of quark--antiquark condensate, or ``zero point'' radiation field in classical sense. Suppose we place a corresponding antiquark in vicinity of the first quark. Owing to their opposite signs, color polarization fields of the quark and antiquark interfere destructively in the overlap region eliminating each other maximally at the midpoint between them  Figure \ref{qaq-field}. This effect leads to a decreasing value of the condensate density in that region and overbalancing the radial vacuum pressure acting on the quark or antiquark. As a result, an attractive force between the quark and antiquark emerges and the quark and antiquark start to move towards each other. This effect is very similar to Casimir effect \cite{casimir}, and we can call it as \emph{color Casimir effect}. We can identify this mechanism of quark-antiquark attraction with  gluon $(R\bar{R},G\bar{G},B
\bar{B})$ exchange between them in QCD.
\begin{figure}[ht]
\centering
\includegraphics[width=0.88\textwidth]{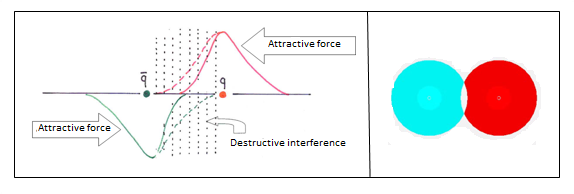}
\vspace{10pt}
\caption{Destructive interference between color fields of quark and anti-quark leads to attractive force between them.}
\label{qaq-field}
\end{figure}
 The density of the resulting condensate around
the quark (antiquark) is identified with the hadronic matter
distribution which is associated with a dynamical mass of the quark. At maximum displacement in the $q\overline{q}$ system
corresponding to small overlap of color fields, hadronic
matter distributions have maximum extent and densities. The quark (antiquark) in this state possesses a constituent mass. The closer
they come each other, the larger is the destructive interference
effect and the smaller hadronic matter distributions around quarks
and the larger their kinetic energies.
\begin{figure}[ht]
\centering
\includegraphics[width=0.88\textwidth]{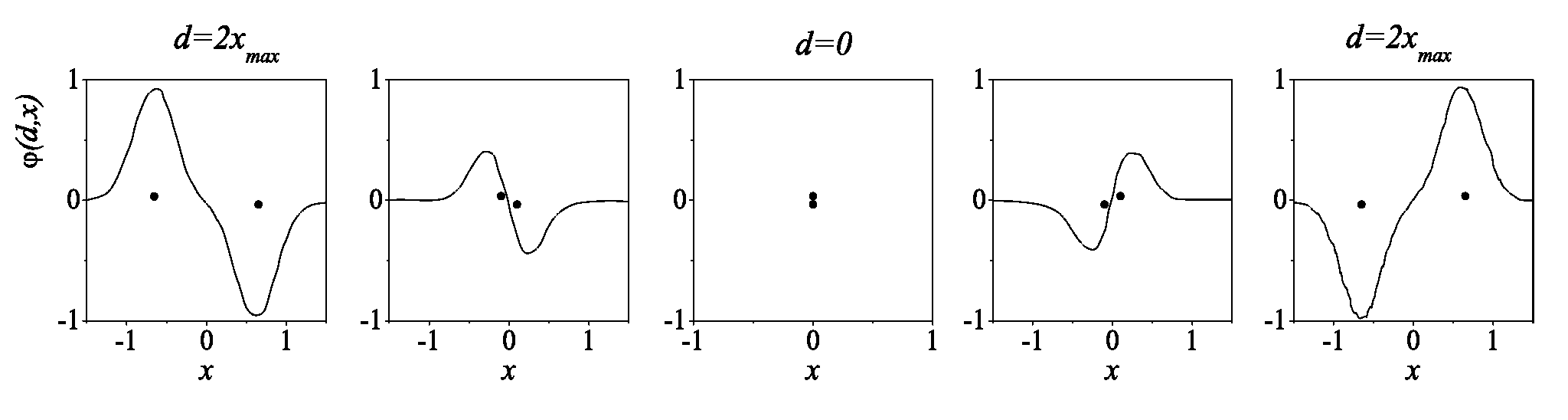}
\caption{Evolution of the color fields/hadronic matter distribution in oscillating quark-aniquark system. Circles depict positions of quark and antiquark at several displacements \emph{d}.}
\label{qaq-oscil}
\end{figure}
 At a minimal displacement the quark (antiquark) becomes a relativistic one with a current mass. So, the quark and antiquark start to oscillate back and forth around their midpoint. Suppose that the profile of color field and density of condensate around the quark (antiquark) has a form close to gaussian. Then oscillation of the quark-antiquark system looks like as shown on Figure \ref{qaq-oscil}. For such interacting $q\overline{q}$ pair located from each other on a
distance $2x$, the total Hamiltonian is
\begin{equation}
H=\frac{m_{q}}{\sqrt{1-\beta ^{2}}}+\frac{m_{\overline{q}}}{\sqrt{1-\beta
^{2}}}+V_{q\overline{q}}(2x),
\end{equation}%
where $m_{q}$, $m_{\overline{q}}$ are the current masses of the valence
 quark and antiquark, $\beta =\beta (x)$ is a velocity depending on their displacement from each other, and $V_{q\overline{q}}$ is the quark--antiquark potential energy at separation $2x.$ It can be rewritten as
\begin{equation}
H=\left[ \frac{m_{q}}{\sqrt{1-\beta ^{2}}}+U(x)\right] +\left[
\frac{m_{\overline{q}}}{\sqrt{1-\beta ^{2}}}+U(x)\right] =H_{q}+H_{\overline{q}},
\label{hamil}
\end{equation}%
where $U(x)=\frac{1}{2}V_{\overline{q}q}(2x)$ is the potential
energy of the quark or antiquark. We postulate that the potential energy of quark is equal to its dynamical mass:
\begin{equation}
U(x)=\int_{-\infty }^{\infty }dz^{\prime }\int_{-\infty }^{\infty
}dy^{\prime }\int_{-\infty }^{\infty }dx^{\prime }\rho (x,{\mathbf{r}%
^{\prime }})\approx M_{Q}(x)
\label{poten-mass}
\end{equation}%
with
\begin{equation}
\rho (x,\mathbf{r}^{\prime })=c\left| \varphi (x,\mathbf{r}^{\prime })\right|
=c\left| \varphi _{Q}(x^{\prime }+x,y^{\prime },z^{\prime })-\varphi _{%
\overline{Q}}(x^{\prime }-x,y^{\prime },z^{\prime })\right|.
\label{color-field}
\end{equation}
where $\rho$ is the resulting density of hadronic matter (quark-antiquark condensate) formed by color fields $\varphi_{Q}$ and $\varphi_{\overline{Q}}$ of the quark and antiquark, respectively. c is a normalization constant. The resulting function $\varphi(d,x)$ shown on Figure \ref{qaq-oscil} represents a potential energy density profile, and its integral (\ref{poten-mass}) corresponds to dynamical mass of quarks. The structure and shape of vacuum polarization around the color quark/antiquark which could give us the information about the confining potential is not known.
It turned out that the mentioned above quark--antiquark system behaves similarly to a so-called {\it breather} solution of one--dimensional Sine-Gordon equation \cite{Raja} which in scaled form reads
\begin{equation}
\Box \phi (x,t)+\sin \phi (x,t)=0,
\end{equation}
where $\phi(x,t)$ is a scalar function and $x$ and $t$ are dimensionless.
The {\it breather} solution
\begin{equation}
\phi(x,t)_{br}=4\arctan \left[(1-\sqrt{(T/2\pi)^{2}}) \frac{\sin(2\pi t/T)}{\cosh(2\pi x/T)} \right]
\label{bre-sol}
\end{equation}
describes the periodic soliton--antisoliton system oscillating with a period $T$ (upper row on Fig. 5).
  The energy density profile of the soliton--antisoliton system
\begin{equation}
\varphi(x,t)_{br}=d\phi(x,t)_{br}/dx
\end{equation}
oscillates (bottom row on Fig. \ref{breather}) the same way as our quark--antiquark system (Fig. \ref{qaq-oscil}).
\begin{figure}[ht]
\centering
\includegraphics[width=0.82\textwidth]{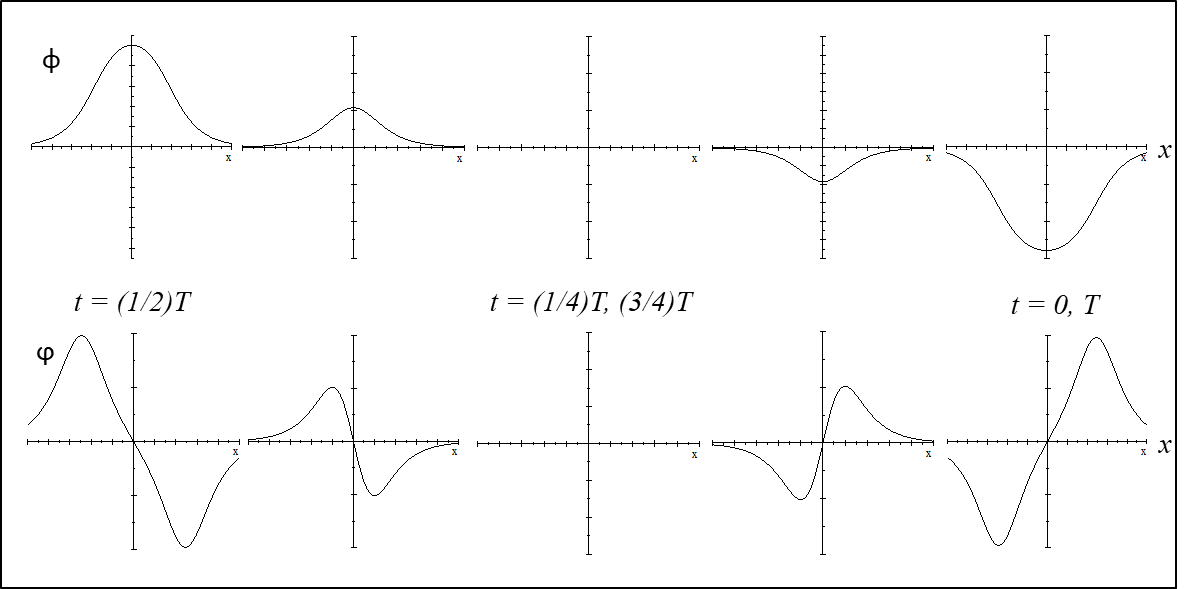}
\caption{Bound soliton-antisoliton or breather, $\phi (x, t)$, (top row) and its energy density profile, $\varphi (x, t)$, (bottom row) at different moments of oscillation with a period $T$.}
\label{breather}
\end{figure}
At maximal displacements ({\it{t = 0,(1/2)T, T}}) the soliton and antisoliton energy density profile, $\phi(x,t)$, is maximal and at minimum displacement ({\it t = (1/4)T, (3/4)T}) they "annihilate". It is not surprising, because our quark--antiquark system is built in close analogy with the model of dislocation--antidislocation \cite{frenkel}, which in its continuous limit is described by breather solution (\ref{bre-sol}) of SG equation.  W. Troost \cite{troost} demonstrated that the Hamiltonian (\ref{hamil}) corresponds to dynamics of the coupled soliton--antisoliton pair with effective potential $U(x)$
\begin{equation}
U(x)=M\tanh ^{2}(\alpha x),
\label{poten-eq}
\end{equation}
where $M$ is a mass of soliton/antisoliton and $\alpha$ is an adjusting parameter. Hence, we can identify our potential of quark--antiquark interaction in hamiltonian (\ref{hamil}) with the potential of soliton--antisoliton interaction.
It is a remarkable fact that a soliton, antisoliton and breather obey
relativistic kinematics, i.e. their energies, momenta and shapes
are transformed according to Lorentz transformations. Since the
above consideration of solitons is purely classical the
important question is how to construct quantum states around them.
Although the soliton solution of SG equation looks like an
extended ``quantum'' particle the relation of classical solitons to
quantum particles is not so trivial. The technique of quantization
of the classical solitons with the usage of various methods has
been developed by many authors. The most known of them is
semiclassical method of quantization (WKB) which allows one to
relate classical periodic orbits (breather solution of SG) with
the quantum energy levels \cite{DHN}.
Another justification of our semiclassical considerations follows from the paper of E. Schr\"{o}dinger \cite{schrod}, where he has shown that wave packet solutions to the time-dependent Schr\"{o}dinger
equation for a harmonic oscillator move in exactly
the same way as corresponding classical oscillators.
These solutions are called ``coherent states''.

Since quarks are the members of the fundamental color triplet,
generalization to the 3-quark system (baryons, composed of Red,
Green and Blue quarks) is performed according to
$SU(3)_{color}$ symmetry: a pair of quarks has coupled
representations $3\otimes 3=6\oplus \overline{3}$ and for quarks
within the same baryon only the $\overline{3}$ (antisymmetric)
representation is realized. Hence, an antiquark can be replaced by
two correspondingly colored quarks to get a color singlet baryon (Figure \ref{aq-2q});
destructive interference takes place between color fields of three
valence quarks (VQs).
\begin{figure}
\centering
\includegraphics[width=0.5\textwidth]{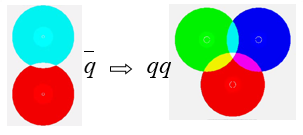}
\caption{Generalization from quark-antiquark to 3-quark system.}
\vspace{-10pt}
\label{aq-2q}
\end{figure}
Destructive interference of quark color fields can be associated with color gluon exchange in QCD. In QCD eight gluons are combined from nine two-color states (\ref{9gluons}), and hamiltonian for gluon exchange between color quarks will be of the form
\begin{equation}\label{inter1}
 H_{12} \sim \mathbf{F_{1}}\cdot \mathbf{F_{2}},
\end{equation}
where $\mathbf{F}=\frac{1}{2}\overline{\lambda}$ are $3\times3$--matrices. For 3 body system the interaction between color quarks has the form
\begin{equation}\label{inter3}
 H_{3q} \sim \sum_{i\neq j}\mathbf{F_{i}}\cdot \mathbf{F_{j}}.
\end{equation}
  In our approach three gluon self--interaction corresponds to destructive interference of R, G, B color fields of three quarks in overlap space near midpoint (white area at the midpoint on Fig.~\ref{3q-c}).
\begin{figure}[ht]
\centering
\includegraphics[width=3.5in]{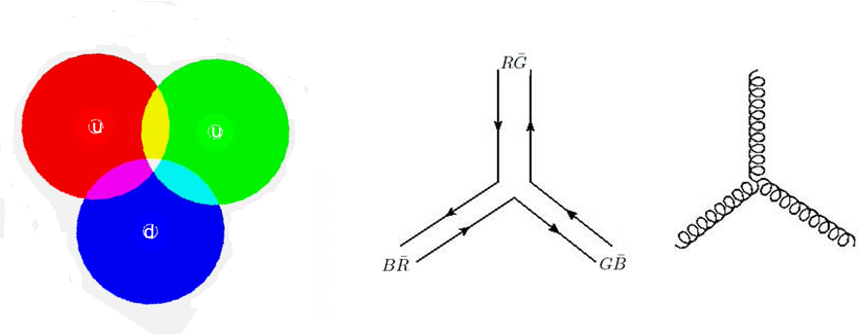}
\caption{Correspondence between overlap of R, G, B color fields of quarks (left) and 3--gluon interaction in QCD (right); possible self-interacting gluon states (middle). Color fields' overlaps between pairs of quarks correspond to gluon exchanges between them.}
\label{3q-c}
\end{figure}
Putting aside the mass and charge differences of valence quarks one can consider three quarks as
oscillating synchronously along the bisectors of equilateral triangle turning from the constituent to current state and inversely. One can see that in this approach 3-body problem arising in treatment of 3-quark dynamics in baryons is solved.
\begin{figure}[ht]
\vspace{-10pt}
\centering
\includegraphics[width=4.7in]{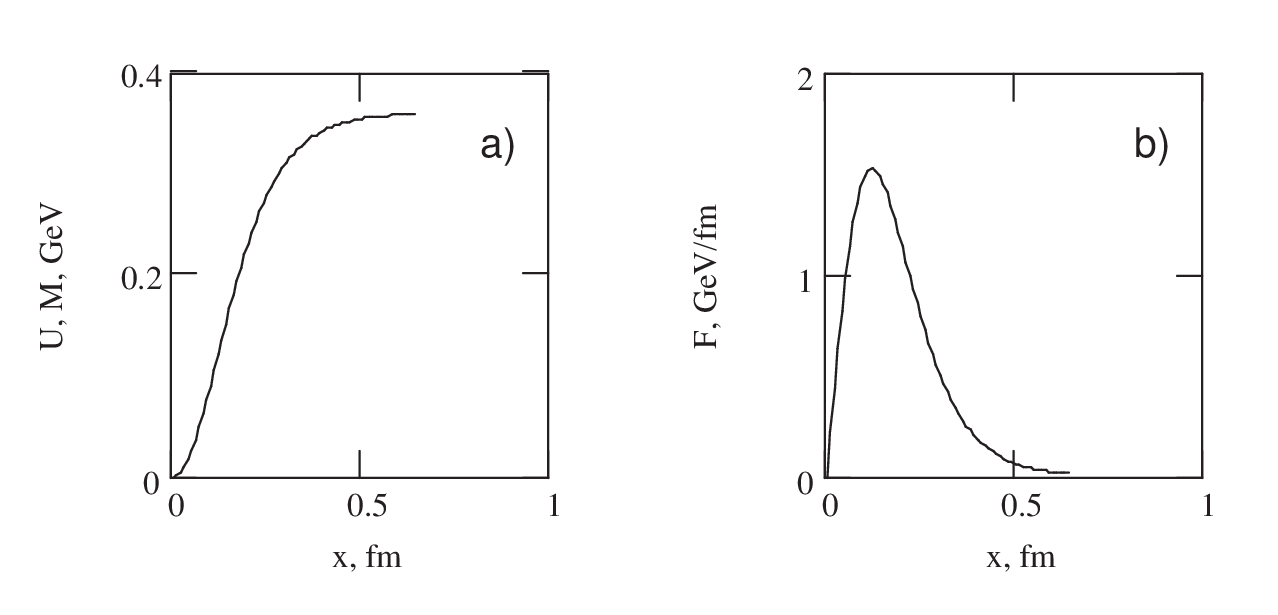}
\vspace{-20pt}
\caption{a) Potential energy of a quark and its dynamical (constituent) mass versus its displacement from the origin of oscillation; b) ``Confinement'' force.}
\label{pot-force}
\vspace{-10pt}
\end{figure}

Therefore, the model unifies the features of relativistic and non-relativistic/constituent models. At a maximal displacement quark becomes non-relativistic with constituent mass corresponding to the maximal value of condensate surrounding it. Further, owing to the prevailing condensate pressure from the outside, it moves under influence of the potential (\ref{poten-eq}) (see Fig. \ref{pot-force}a) towards two other quarks, and at the origin of oscillation it becomes relativistic with the current mass. Thus, during oscillation quarks transit from constituent states to current states and inversely, that corresponds to \textit{dynamical} restoration/braking of chiral symmetry. A nucleon composed of three oscillating quarks runs over the states corresponding to terms of the infinite series of Fock space:
\begin{equation}\label{fock}
 |B\rangle = c_{1}|q_{1}q_{2}q_{3}\rangle + c_{2}|q_{1}q_{2}q_{3}\bar{q}q\rangle + ...
\end{equation}
\begin{figure}[ht]
\vspace{-10pt}
\centering
\includegraphics[width=4.7in]{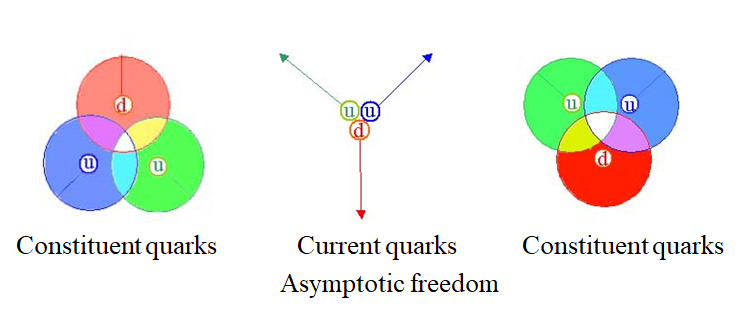}
\vspace{-10pt}
\caption{Interplay between constituent and current quark states inside a baryon. Color fields and/or hadronic matter distributions of 3-quark system (baryon) are depicted as \textit{Red, Green, Blue} circles at different moments of oscillation corresponding to minimal and maximal quarks displacements, $d=0, 2x$.}
\label{cur-const}
\end{figure}
Important feature of the model is that there is no a confining potential/force inside a nucleon.
During oscillations (putting aside Coulomb and spin interactions) the interaction force between quarks $F(x) = -dU/dx$ vanishes while quarks are at both the origin of oscillation and maximal displacement (Fig. \ref{pot-force}b). It becomes maximal in between the origin and maximal displacement. Thus, at the origin of oscillations, where the quark and antiquark in mesons and three quarks in baryons being as relativistic ones do not interact, the state of asymptotic freedom is realized (Fig. \ref{cur-const}).
\begin{figure}[tb]
\vspace{-10pt}
\begin{center}
\includegraphics[width=3.5in]{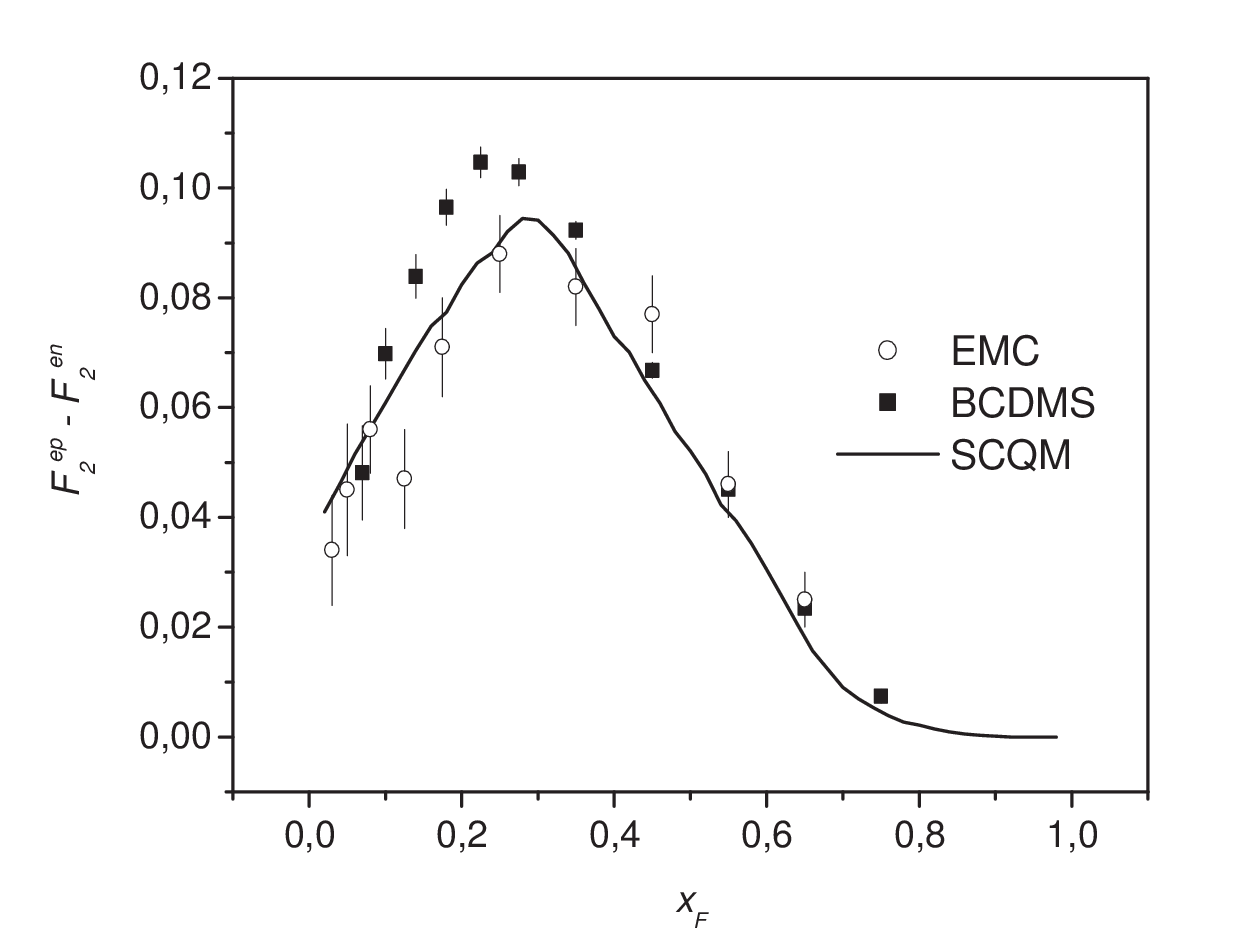}
\end{center}
\vspace{-10pt}
\caption{Valence quark structure functions in nucleons; data are from
papers \cite{emc}, \cite{bcdms}.}%
\label{emc}
\end{figure}
The model reproduces the momentum distribution of valence quarks in a nucleon measured by collaborations EMC \cite{emc} and BCDMS \cite{bcdms} in deep inelastic lepton scattering (DIS) experiments (Figure \ref{emc}). They measured structure functions $F_{2}^{ep}$ and $F_{2}^{en}$ for the proton and neutron, correspondingly, and their difference
\begin{equation}
F_{2}^{ep}-F_{2}^{en}=\frac{x}{3}\left[  u_{v}(x)-d_{v}(x)\right],
\end{equation}
that relates to the momentum distribution of the valence quarks inside the nucleon. Here $x$ is a fraction of nucleon momentum carried by a quark and $u_{v}$ and $d_{v}$ are density distributions of quarks. In the model they are determined by equations of motion
\begin{equation}
\dot{p}_q = \frac{\partial H_{q}}{\partial x} ~ \textrm{and} ~ \dot{x}_q = \frac{\partial H_{q}}{\partial p_q},
\label{eqs_motion}
\end{equation}
where hamiltonian $H_{q}$ is defined by Eq. (\ref{hamil}).

The model meets local gauge invariance. Indeed, suppose $\psi_{color}$ is a wave function of a single quark in color space where index {\it color} corresponds to one of the values Red, Green, Blue. Interactions of R, G, and B quarks in a nucleon which result in their oscillations can be reduced to the phase rotation of the wave function $\psi_{color}$ of each quark in its color space
\begin{equation}
\psi_{color}(x) \rightarrow e^{i\theta(x)}\psi_{color}(x).
\label{psi_col}
\end{equation}
This phase rotation results in dressing (undressing) of the quark by quark--antiquark condensate that is linked with transformation of the gauge field $A^{\mu}_{a}$:
\begin{equation}
A^{\mu}_{a}(x) \rightarrow A^{\mu}_{a}(x)+\partial^{\mu}\theta_{a}(x),
\label{gauge-eq}
\end{equation}
where $a$ is a color index.
Such consideration comes from correlated dynamics of quark and antiquark and three quarks in mesons and baryons accordingly.
As 3-quark system (baryon) has been constructed on that fact that any pair of quarks is a member of antitriplet $\overline{3}$ we can simplify our consideration of baryons by the quark-antiquark system which is a color singlet
\begin{equation}\label{singlet}
\sqrt{\frac{1}{3}}(R\overline{R}+G\overline{G}+B\overline{B}).
\end{equation}
 Strong correlation of quarks coming from destructive interference of opposite color fields of quark and antiquark allows us to take into consideration only a single colored quark (one of $R, G, B$) described by a single component function (\ref{psi_col}). At maximal displacement, where the phase $\theta = 0, \pi, 2\pi$,  the color field of quark (Fig. \ref{qaq-oscil}) and the wave function $\psi_{color}$ have a maximal value; at the midpoint of oscillation, where the phase $\theta = (1/2)\pi, (3/2)\pi$, the quark color field and $\psi_{color}$  go to zero (due to destructive interference). Therefore, the strong interaction between quarks (Eqs. (\ref{inter1}) and (\ref{inter3})), described in our model by (\ref{hamil}), is effectively reduced to the phase rotation of the quark wave function in a single color space that, in turn, allows one to drop the color indices of a gauge field: $A^{\mu}_{a}(x) \rightarrow A^{\mu}(x)$.
Hence interaction of color quarks via non--Abelian fields of QCD in our model is ``reduced'' to its electrodynamical analog
\begin{equation}
F^{\mu\nu}_{a}=\partial^{\mu}A^{\nu}_{a}-\partial^{\nu}A^{\mu}_{a}-\lambda f^{abc}A^{\mu}_{b}A^{\nu}_{c} \rightarrow F^{\mu\nu}=\partial^{\mu}A^{\nu}-\partial^{\nu}A^{\mu}.
\label{Field}
\end{equation}
Since destructive interference is maximal in overlapping region of color fields of quark and antiquark in mesons and three quarks in nucleons, mesons and nucleons are colored objects. I.e. a nucleon is colorless in small area at its center of mass, but becomes \emph{three colored} (R,G,B) on periphery. This feature turns out to be important when nucleons arrange inside nuclei lattice-like structure.

What concerns quark confinement, apparently, ``imprisonment'' of quarks is a consequence of the topological nature of hadrons. Since we elucidated our model in analogy with the breather which is a topological object, we assume that the quark--antiquark describing mesons and three quark systems describing baryons are topological solitons. Topological solitons are characterized by the conserving numbers, so--called, winding numbers. For baryons the winding number is identified with the baryon number.  It means that at any temperature and density of nuclear environment baryon conserves its identity, i.e. baryonic number. Considerations on the spin content of nucleon in the framework of SCQM is given in \cite{Mus4}.

The parameters of the model are the maximum displacement of valence quark and antiquark in mesons and 3 quarks in baryons, $x_{\max }$, and dimension of the hadronic matter distribution formed by quark--antiquark condensate around them. In absence of knowledge about the shape of quark--antiquark condensate around valence quarks, or the form of hadronic matter in a constituent quark $\varphi_{Q(\overline{Q})}$, we take it in a gaussian form
\begin{equation}
\varphi _{Q(\overline{Q})}(x,y,z)=\varphi _{Q(\overline{Q})}(x_{1},x_{2},x_{3})=%
\frac{(\det \hat{A})^{1/2}}{(\pi )^{3/2}}\exp \left( -\mathbf{X}^{T}\hat{A}%
\mathbf{X}\right) ,
\label{gaussian}
\end{equation}
where the exponent is written in a quadratic form. Here $\mathbf{X} = {x_1, x_2, x_3}$ and $\hat{A}$ is the matrix of quadratic form. We set $z$ axis perpendicular to the plane of quark oscillations.
The value of maximal quark displacement, and parameters of the gaussian form of hadronic matter distribution around VQs are estimated with the usage of, so-called, overlap function  applied to analyse $pp$ and $\overline{p}p$ collisions \cite{Mus1}.

\subsection{Adjustment of Parameters}
With the aim to derive the parameters of the model we perform Monte-carlo simulation of nucleon-nucleon collisions at high energies. Different configurations of quark content inside a nucleon  (Figure \ref{cur-const})
realized at the instant of the collision result in different types
of reactions. Configurations with non-relativistic constituent quarks ( $x_{disp} \simeq
x_{max}$) in both colliding nucleons lead to soft
interactions with nondiffractive multiparticle production in
central and fragmentation regions (Fig. \ref{q-config}a). Hard
scattering with jet production and large angle elastic
scattering take place when configurations with current VQs
($x_{disp} \simeq0$) in both colliding nucleons are realized (Fig.
\ref{q-config}b).
The near current quark configuration inside one
of the nucleons and constituent quark configuration inside the
second one result in single diffraction scattering (Fig.
\ref{q-config}c). And at last intermediate configurations inside
one or both nucleons are responsible for semihard and double
diffractive scattering (Fig. \ref{q-config}d). The same
geometrical consideration can be applied to deep inelastic
scattering processes if one assumes that a real or virtual photon
converts into the vector meson according to the vector dominance
model.
\begin{figure}
\vspace{-20pt}
\centering
\includegraphics[width=4.2in]{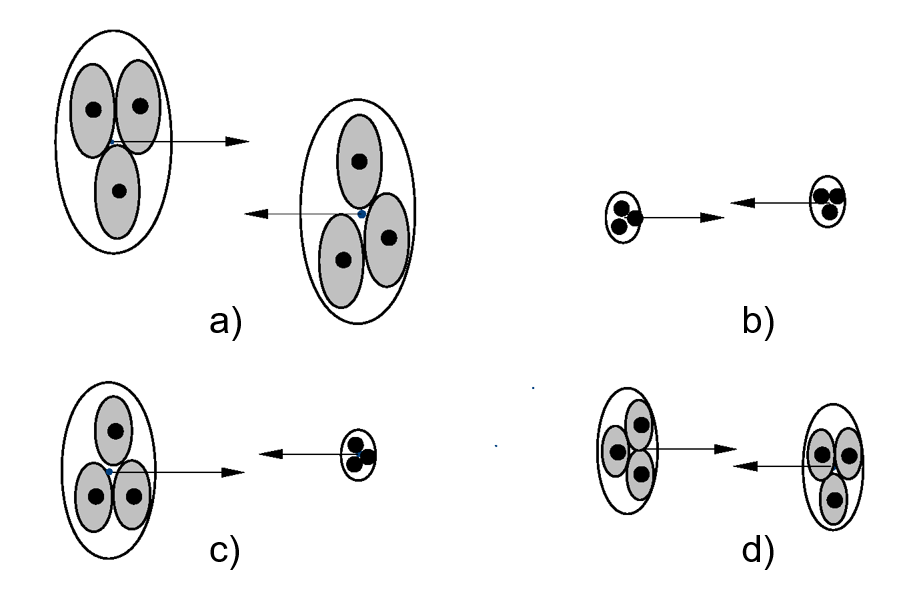}
\vspace{-20pt}
\caption{Different quark configurations realized inside colliding nucleons
at
the instant of collision.}%
\label{q-config}%
\end{figure}

Using impact parameter representation,
namely Inelastic Overlap Function (IOF), we can calculate total, inelastic,
elastic and single diffractive cross sections for $pp$ and $\overline{p}p$
collisions. In impact parameter representation IOF can be specified via the
unitarity equation
\begin{equation}
2Imf(s,b)=|f(s,b)|^{2}+G_{in}(s,b),
\end{equation}
where $f(s,b)-$elastic scattering amplitude and $G_{in}(s,b)$ is IOF. IOF can be expressed via eikonal
\begin{equation}
G_{in}(s,b)=1-\exp^{-2Im\chi(s,b)},
\end{equation}
where $\chi(s,b)$ is eikonal, that given by the Fourier-Bessel transform of the square of the proton form factor. IOF is
connected with inelastic differential cross sections in impact parameter space:
\begin{equation}
\frac{1}{\pi}(d\sigma_{in}/db^{2})=G_{in}(s,b).
\label{iof}
\end{equation}
\begin{figure}[tb]
\vspace{-20pt}
\begin{center}
\includegraphics[width=2.2433in]{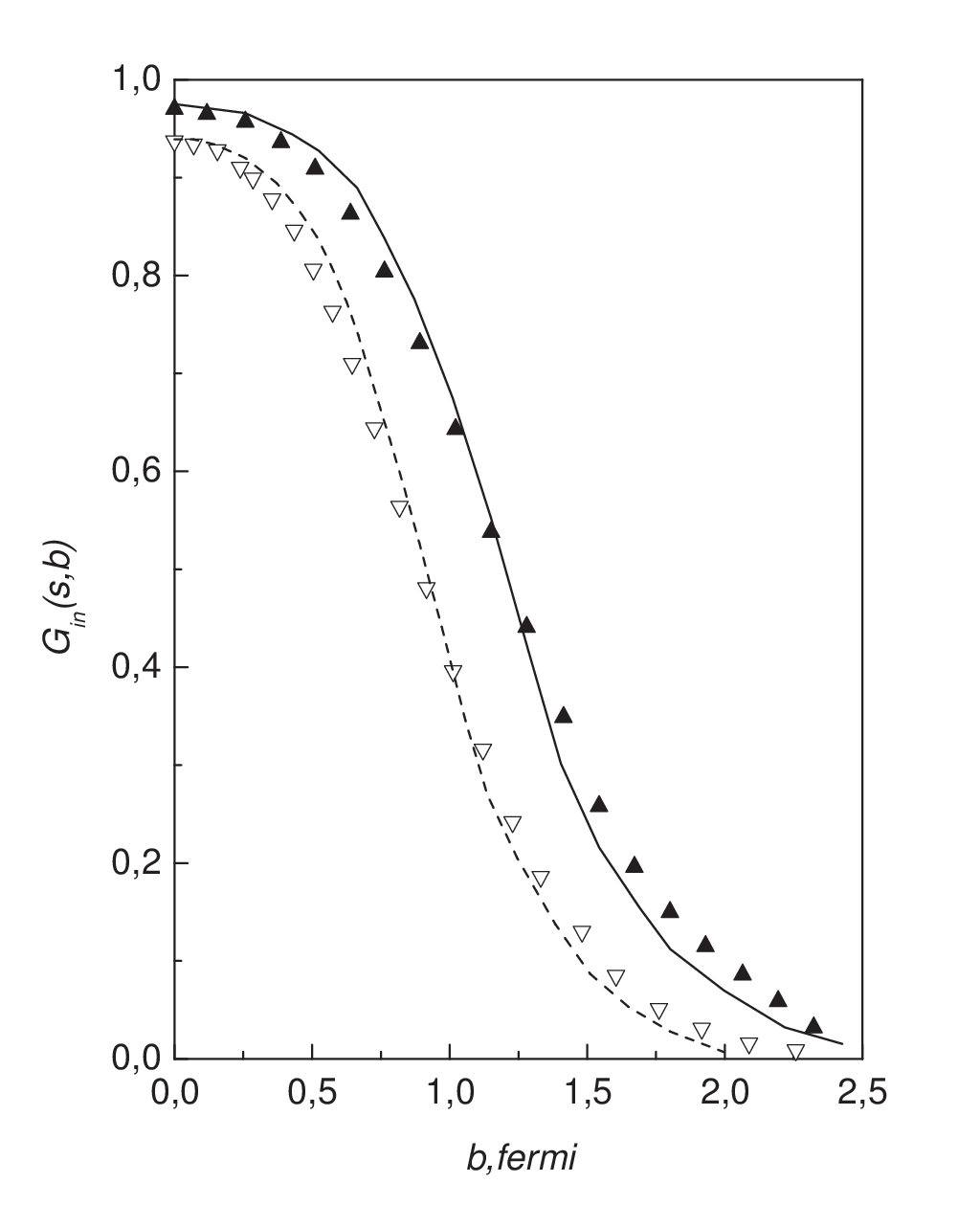}
\end{center}
\vspace{-20pt}
\caption{Curves are calculated IOPs for $pp$ and $\overline{p}p-$ collisions
at $\sqrt{s}=53$ (dashed) and $540$ $GeV$ (solid); triangles are results of Henzi and Valin
parametrization of experimental data \cite{henzi}.}%
\label{henzi}
\end{figure}
Then inelastic, elastic and total cross sections can be expressed
via IOF as%

\begin{equation}
\sigma_{in}(s)=\int G_{in}(s,\mathbf{b})d^{2}\mathbf{b},
\label{sin}
\end{equation}%

\begin{equation}
\sigma_{el}(s)=\int\left[  1-\sqrt{1-G_{in}(s,\mathbf{b})}\right]  ^{2}%
d^{2}\mathbf{b},
\label{sel}
\end{equation}%
\begin{equation}
\sigma_{tot}(s)=2\int\left[  1-\sqrt{1-G_{in}(s,\mathbf{b})}\right]
d^{2}\mathbf{b}.
\label{stot}
\end{equation}
\begin{figure}[tb]
\vspace{-20pt}
\begin{center}
\includegraphics[width=4.7in]{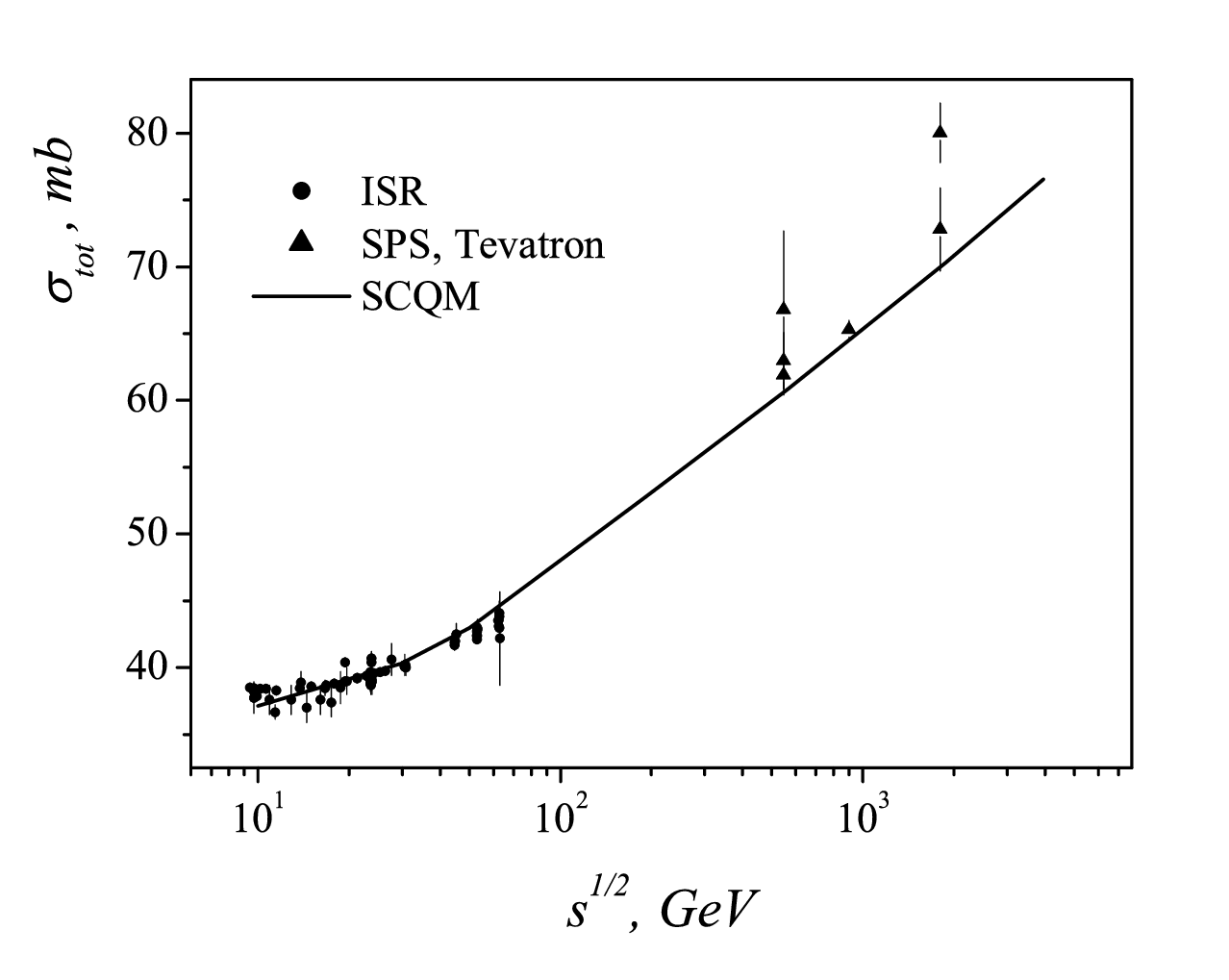}
\end{center}
\vspace{-20pt}
\caption{Total cross section for $pp$ and $\overline{p}p-$ collisions; data
points are compilations of experimental data taken from electronic data base
HEPDATA\cite{hep}.}
\label{tot-cs}
\end{figure}

Since IOF relates to the probability of inelastic interaction at
given impact parameter, (\ref{iof}), we carried out Monte Carlo simulation of
nucleon--nucleon interactions. Inelastic interaction takes place at
certain impact parameter $b$ if at least one pion is produced in the region
where hadronic matter distributions of colliding protons overlap
\begin{equation}
4M_{p_{i}}\gamma_{p_{i}}M_{t_{j}}\gamma_{t_{j}}\int\rho_{p_{i}}(\mathbf{r}%
)\rho_{t_{j}}(\mathbf{r}-\mathbf{r}^{\prime})~d^{3}\mathbf{r}\geq
m_{\pi_{\perp}}^{2},
\end{equation}
where indices $q_{i}$ and $p_{j}$ refer to quarks from projectile and target nucleons correspondingly, and
$i,j=1,2,3$, $M_{p_{i}}$, $M_{t_{j}}$ -- masses of hadronic matter composed in
constituent quarks $p_{i}$ and $t_{j}$, $\gamma_{q_{i}}$, $\gamma_{p_{j}}$ --
their $\gamma$-factors; intergrand expression is convolution of hadronic
matter density distributions of quarks $q_{i}$ and $p_{j}.$ This condition
corresponds Hisenberg picture \cite{heisen} with modified right hand side:
$m_{\pi}^{2}$ in the original Heisenberg inequality is replaced by
$m_{\pi_{\perp}}^{2}.$ It is justified by the fact that the average transverse
momentum of produced particles increases with energy.
\begin{figure}
\begin{center}
\includegraphics[width=4.5in]{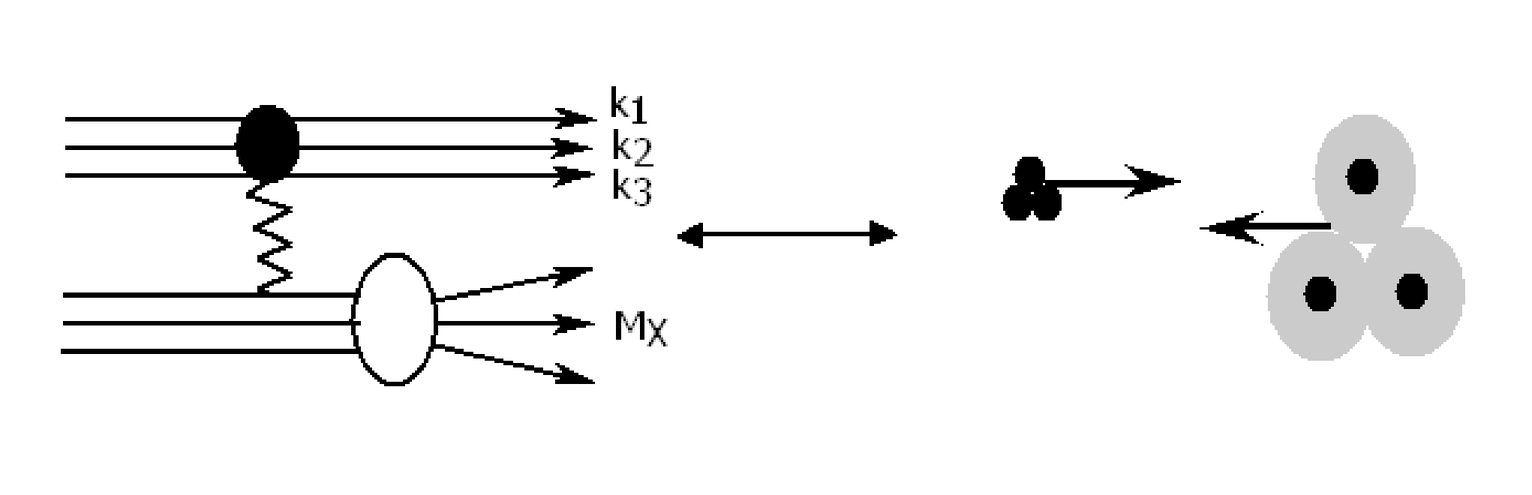}
\end{center}
\vspace{-25pt}
\caption{Correspondence between diagram and geometrical representation of
single diffractive dissociation in $pp\rightarrow pX$ processes. Large grey
circles are constituent quarks of dissociating proton, small black circles --
current quarks of the another proton.}%
\vspace{-15pt}
\label{SD-diag}
\end{figure}
\begin{figure}
\vspace{-35pt}
\begin{center}
\includegraphics[width=4.5in]{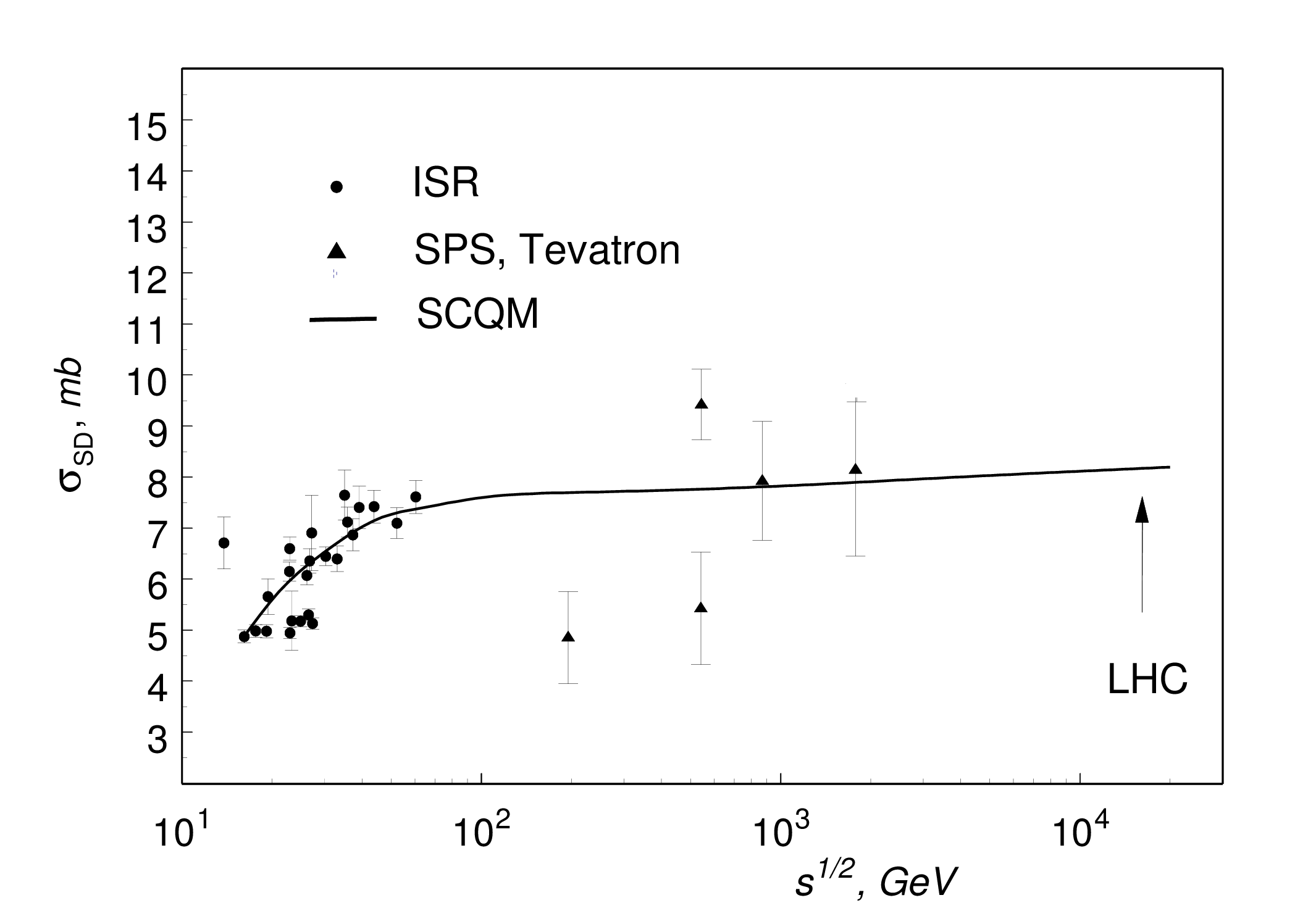}
\end{center}
\vspace{-20pt}
\caption{Single diffractive dissociation cross sections for $pp$ and
$\overline{p}p-$ collisions; data points are compilation of experimental data
taken from electronic data base HEPDATA\cite{hep}.}
\label{SD-cs}
\end{figure}
Therefore, we determine
$G_{in}(s,b)$ for certain values of impact parameter $b$ according to formula (\ref{iof}) that allows us to calculate cross
sections $\sigma_{in}, \sigma_{el}$ and $\sigma_{tot}$ by expressions (\ref{sin}), (\ref{sel}) and (\ref{stot}). The values of adjusted
parameters of the model, $\alpha$ and the maximal displacement of quarks, $x = x_{\max}$ in Eq. (\ref{poten-eq}), parameters of qaussian distribution $\sigma_{x,y}$ and $\sigma_{z}$ in Eq. (\ref{gaussian}) chosen by comparison of calculated IOF with so
called ``BEL''--parametrization of experimental data, (Figure \ref{henzi}), are given in Table \ref{tbl1}. These values correspond to the root mean square radius of a proton 0.85$\pm${0.03} fm that is in a good agreement with the experimental value 0.84 fm.  The mass of the constituent quark at maximum displacement is taken as $%
M_{Q(\overline{Q})}(x_{\max })=\frac{1}{3}\left( \frac{m_{\Delta }+m_{N}}{2}%
\right)$, where $m_{\Delta }$ and $m_{N}$ are
masses of the delta isobar and nucleon correspondingly. The current
mass of the valence quark is taken to be $5\ ${MeV}. All parameters of the model are given in Table \ref{tbl1}.
\begin{table}[!ht]
\caption{Model Parameters }
\vspace*{6pt}
\begin{center}
{\begin{tabular}{@{}cccccc@{}}
\hline
  {$m_q, MeV$} & {$M_Q, MeV$} & $\alpha$ & {$x_{max}, fm$} & {$\sigma_{x,y}, fm$} & {$\sigma_{z}, fm$}\\
\hline
  5 & 360 & 5.5 & 0.64$\pm${0.02} & 0.24$\pm${0.01} & 0.12$\pm${0.01}\\
\hline
\end{tabular}
}
\end{center}
\label{tbl1}
\end{table}
Energetic dependence of
total cross sections compared with experimental ones, $\sigma_{in}(s),$ in $pp$ and
$\overline{p}p-$ collisions at wide range of energy are shown on Figure \ref{tot-cs}.
Oscillatory motion of VQs appearing as interplay between constituent and
bare (current) quark configurations results in fluctuations of hadronic matter
distribution inside colliding nucleons. The manifestation of these
fluctuations is a variety of scattering processes, hard and soft, in
particular, the process of single diffraction (SD). We select SD--events among
inelastic $pp\rightarrow pX$ events with the criterion $1-x_{F}<0.1$, where
$x_{F}=\frac{2}{\sqrt{s}}(k_{1}+k_{2}+k_{3})$ (Figure \ref{SD-diag}). Here $k_{1},$ $k_{2},$
$k_{3}$ are momenta of quarks forming the final state proton. As one can see,
SD--events correspond to constituent quark configuration inside one colliding
proton and (semi)bare quark configuration inside another one. Figure \ref{SD-cs} shows
that calculated SD cross section slightly depends on energy; our calculation
for LHC energy gives $\sigma_{SD}^{LHC}(pp)=8.30\pm0.15$ $mb.$ This result has been obtained long before recent measurements of SD cross section at LHC which differs from ~ 7 mb through 11 mb given by different experiments. Our prediction is in a good agreement with the result of ATLAS Collaboration: $SD_{Atlas}=7.9 mb$ \cite{ATLAS}.

 \section*{Conclusions}
Proposed the semi-empirical quark model of nucleon structure SCQM possesses the features of both non-relativistic and relativistic quark models. Based on SU(3) color symmetry it includes the main features of QCD: local gauge invariance, asymptotic freedom, and chiral symmetry breaking.
In view of correlated plane oscillation of valence quarks and non-spherical hadronic matter distribution around the valence quarks, we claim that the nucleon is \emph{non-spherical, oblate} object. As will be shown in the Second Part, the oblate triangular (with color round corners) shape of nucleon becomes apparent in peculiarity of nuclear density distributions and the sizes of nuclei. Namely oblate triangular nucleons can form and make understandable the variety of nuclear structures. These structures appear as quark-quark correlations between nucleons which, in turn, lead to nucleon-nucleon correlations.

\end{document}